\documentclass[a4paper]{jpconf}

\usepackage{graphicx}

\usepackage{cite}
\usepackage{physics}

\usepackage{tikz}
\usetikzlibrary{arrows,shapes}
\usetikzlibrary{trees}
\usetikzlibrary{matrix,arrows} 				% For commutative diagram
\usetikzlibrary{positioning}				% For "above of=" commands
\usetikzlibrary{calc,through}				% For coordinates
\usetikzlibrary{decorations.pathreplacing}  % For curly braces
\usepackage{pgffor}							% For repeating patterns
\usetikzlibrary{decorations.pathmorphing}	% For Feynman Diagrams
\usetikzlibrary{decorations.markings}
\tikzset{
	vector/.style={decorate, decoration={snake}, draw},
	provector/.style={decorate, decoration={snake,amplitude=2.5pt}, draw},
	antivector/.style={decorate, decoration={snake,amplitude=-2.5pt}, draw},
	fermion/.style={draw=black, postaction={decorate},
	decoration={markings,mark=at position .55 with {\arrow[draw=black]{>}}}},
	fermionbar/.style={draw=black, postaction={decorate},
	decoration={markings,mark=at position .55 with {\arrow[draw=black]{<}}}},
	fermionnoarrow/.style={draw=black},
	gluon/.style={decorate, draw=black,
	decoration={coil,amplitude=4pt, segment length=5pt}},
	scalar/.style={dashed,draw=black, postaction={decorate},
	decoration={markings,mark=at position .55 with {\arrow[draw=black]{>}}}},
	scalarbar/.style={dashed,draw=black, postaction={decorate},
	decoration={markings,mark=at position .55 with {\arrow[draw=black]{<}}}},
	scalarnoarrow/.style={dashed,draw=black},
	electron/.style={draw=black, postaction={decorate},
	decoration={markings,mark=at position .55 with {\arrow[draw=black]{>}}}},
	bigvector/.style={decorate, decoration={snake,amplitude=4pt}, draw},
}

\def\dm{\Delta m^2}

\newcommand{\asym}[1]{\mathcal{A}^\mathrm{#1}}

\newcommand{\comp}[1]{A^{\mathrm{#1}}}

\begin{document}
\title{Disentangling genuine from matter-induced CP violation in neutrino oscillations }

\author{A Segarra and  J Bernabeu}

\address{Departament de Física Teòrica \& IFIC, Universitat de València - CSIC, 
C/Dr. Moliner 50, E-46100 Burjassot (Spain)}

\ead{alejandro.segarra@uv.es, jose.bernabeu@uv.es}

\begin{abstract}
	We prove~\cite{Bernabeu:2018twl} that, in any flavor transition, 
	neutrino oscillation CP violating asymmetries in matter have two disentangled components: 
	(a) a CPT-odd T-invariant term, non-vanishing iff there are interactions with matter; 
	(b) a T-odd CPT-invariant term, non-vanishing iff there is genuine CP violation. 
	As function of the baseline, these two terms are distinct $L$-even and $L$-odd observables, respectively. 
	In the experimental region of terrestrial accelerator neutrinos, 
	we calculate~\cite{Bernabeu:2018use} their approximate expressions from which we prove that, 
	at medium baselines, the CPT-odd component is small and nearly $\delta$-independent, 
	so it can be subtracted from the experimental CP asymmetry as a theoretical background, 
	provided the hierarchy is known. 
	At long baselines, on the other hand, we find that 
	(i) a Hierarchy-odd term in the CPT-odd component dominates the CP asymmetry 
	for energies above the first oscillation node, 
	and (ii) the CPT-odd term vanishes, independent of the CP phase $\delta$, 
	at $E=0.92~\mathrm{GeV} (L/1300~\mathrm{km})$ near the second oscillation maximum, 
	where the T-odd term is almost maximal and proportional to $\sin\delta$. 
	A measurement of the CP asymmetry in these energy regions would thus provide separate information on 
	(i) the neutrino mass ordering, 
	and (ii) direct evidence of genuine CP violation in the lepton sector.
\end{abstract}

\section{Introduction}

%Over the last two decades,
%experimental studies of flavor oscillations in the
%atmospheric~\cite{Fukuda:1998mi}, solar~\cite{Ahmad:2002jz},
%reactor~\cite{Araki:2004mb} and accelerator~\cite{Ahn:2006zza}
%neutrino sectors have lead to a consistent picture of neutrino oscillations.
Neutrino oscillations arise from the mismatch between the three
%This phenomenon arises from the mismatch between the three
interaction (flavor) eigenstates and the propagation (mass) eigenstates,
described by the PMNS mixing matrix, $\nu_\alpha = \sum_k U_{\alpha k} \nu_k$.
In the standard parametrization, this matrix is written as
\begin{equation}
	U = 		
        \mqty[
                1 &0 &0\\
		0 &c_{23} &s_{23}\\
		0 &-s_{23} &c_{23}
        ]
        \mqty[
		c_{13} &0 &s_{13}\, e^{-i\delta}\\
                0 &1 &0\\
		-s_{13}\, e^{i\delta} &0 &c_{13}
        ]
        \mqty[
		c_{12} &s_{12} &0\\
		-s_{12} &c_{12} &0\\
                0 &0 &1
        ]\,,
\end{equation}
in terms of three mixing angles 
$(c_{ij} = \cos\theta_{ij},\, s_{ij} = \sin\theta_{ij})$
and a CP-violating phase $\delta$.
Its effects in the flavor evolution Hamiltonian
\begin{equation}
    H =		
        \frac{1}{2E}\,
                U
        \mqty[
                m_1^2 &0 &0\\
                0 &m_2^2 &0\\
                0 &0 &m_3^2
        ]
        U^\dagger
\end{equation}
lead to the appearance of energy ($E$) and baseline ($L$) dependent 
oscillations with probability
\begin{equation}
	\label{eq:Prob}
	P_{\alpha\beta} \equiv 
	P(\nu_\alpha \to \nu_\beta)
	= \, \delta_{\alpha\beta}
	-4\sum_{j<i}\mathrm{Re}~J_{\alpha\beta}^{ij}\,
	\sin^2 \Delta_{ij}
	-2\sum_{j<i}\mathrm{Im}~J_{\alpha\beta}^{ij}\,
	\sin 2\Delta_{ij}\,,
\end{equation}
where 
$J_{\alpha\beta}^{ij} \equiv U_{\alpha i} U^*_{\alpha j} U^*_{\beta i} U_{\beta j}$
are the rephasing-invariant mixings,
$\Delta_{ij} \equiv \frac{\Delta m^2_{ij} L}{4 E}$
are the oscillation phases, and
$\dm_{ij} \equiv m^2_i - m^2_j$
are the neutrino mass-squared differences.

Even though the absolute scale of the neutrino masses is still unknown,
both mass differences $\dm_{21}$ and $\abs{\dm_{31}}$,
as well as all three mixing angles,
have reached the precision era~\cite{deSalas:2017kay},
\begin{equation}
	\begin{aligned}
		\dm_{21} &= 7.55(20) \times 10^{-5}~\mathrm{eV}^2\\
		\abs{\dm_{31}} &= 2.50(3) \times 10^{-3}~\mathrm{eV}^2
	\end{aligned}
	\hspace{2cm}
	\begin{aligned}
		&s_{12}^2 = 3.20(20) \times 10^{-1}\\
		&s_{23}^2 = 5.51(30) \times 10^{-1}\\
		&s_{13}^2 = 2.160(83) \times 10^{-2}
	\end{aligned}
\end{equation}
The main goals of the next generation experiments,
such as DUNE~\cite{Acciarri:2015uup} and T2HK~\cite{Hyper-Kamiokande:2016dsw},
will be the measurement of the CP phase $\delta$
and the $\mathrm{sign}(\dm_{31})$, which will determine whether
the mass Hierarchy is Normal ($m_1 < m_2 < m_3$)
or Inverted ($m_3 < m_1 < m_2$).

Notice that there are terms in the oscillation probability~(\ref{eq:Prob})
that are $\delta$-dependent through the CP-conserving $\cos\delta$,
coming from $\mathrm{Re}~J_{\alpha\beta}^{ij}$.
Although they allow to extract the value of the parameter $\delta$,
their measurement cannot be considered as observation of CP violation.
In this work, we study the feasibility of a direct observation
of CP violation in the lepton sector,
in as much a model-independent manner as possible.
Such a probe must come from the measurement of 
a non-vanishing value of a CP-odd observable like the asymmetry $\asym{CP}\equiv P(\nu) - P(\bar\nu)$.
The complication in this measurement stems from the propagation of neutrinos through the Earth.
Since matter is CP asymmetric, it induces a \emph{fake} contribution to the CP asymmetry,
contaminating the test of CP via $\asym{CP}$.

In the next Section,
we exploit the different behavior of the different terms in the oscillation probability
under the discrete symmetries CP, T and CPT
to cleanly separate genuine from matter-induced terms in the CP asymmetry.
This disentanglement will lead to peculiar dependencies of the separate components of $\asym{CP}$
in the mixing parameters.
In Section~\ref{sec:signatures},
we show the signatures induced at future accelerator experiments
by these ideas, focusing especially on the determination of the Hierarchy
and the direct observation of CP violation.
Our conclusions are presented in Section~\ref{sec:conclusions}.

\section{Asymmetry Disentanglement Theorem}
\label{sec:theorem}

The matter effects in neutrino oscillations are described by the
Hamiltonian~\cite{Wolfenstein:1977ue, Barger:1980tf, Kuo:1989qe, Zaglauer:1988gz, Krastev:1990gz, Bernstein:1991ss}
\begin{equation}
	\label{eq:Htilde}
    H =		
        \frac{1}{2E}
        \left\{
                U
        \mqty[
                m_1^2 &0 &0\\
                0 &m_2^2 &0\\
                0 &0 &m_3^2
        ]
        U^\dagger
		+
        \mqty[
                \,a\, &\,0\, &\,0\,\\
                0 &0 &0\\
                0 &0 &0
        ] \right\}
        =\frac{1}{2E}\; \tilde U \tilde M^2 \tilde U^\dagger\,,
\end{equation}
where $a = 2EV$ is the energy-dependent matter parameter,
proportional to the matter potential $V$.
The same Hamiltonian applies to antineutrinos changing 
$\delta \mapsto -\delta$ in $U$, and $a\mapsto -a$.
In practice, the same analytical expressions as in vacuum can be used
to describe neutrino oscillations in matter, 
if one writes them in terms of the energy-dependent 
masses $\tilde M^2$ and mixings $\tilde U$ obtained
from the diagonalization of the Hamiltonian~(\ref{eq:Htilde}).

In general, let's assume there are different masses and mixings for
neutrinos ($\tilde M^2,\, \tilde U$) and 
antineutrinos ($\tilde{\bar M}^2,\, \tilde{\bar U}$),
given by an arbitrary number (at least 3) of eigenstates.
Using the probability in Eq.~(\ref{eq:Prob}) for both of them
we can compute the CP asymmetry,
\begin{align}
	\asym{CP}_{\alpha\beta} &\equiv
	P(\nu_\alpha \to \nu_\beta) - P(\bar\nu_\alpha \to \bar\nu_\beta) =\\
	\nonumber
	&=
	-4\sum_{j<i}\left[
		\mathrm{Re}~\tilde J^{ij}_{\alpha\beta} \sin^2\tilde\Delta_{ij}
		-\mathrm{Re}~\tilde {\bar J}^{ij}_{\alpha\beta} \sin^2\tilde{\bar \Delta}_{ij}
	\right]
	-2\sum_{j<i}\left[
		\mathrm{Im}~\tilde J^{ij}_{\alpha\beta} \sin 2\tilde\Delta_{ij}
		-\mathrm{Im}~\tilde {\bar J}^{ij}_{\alpha\beta} \sin 2\tilde{\bar \Delta}_{ij}
	\right]\,.
\end{align}

In the CPT-invariant limit, as happens in vacuum,
the firs term of the CP asymmetry vanishes due to
$\tilde {\bar M}^2 = \tilde M^2$ and $\tilde{\bar U} = \tilde U^*$.
Conversely, the second one vanishes in the T-invariant limit,
since the real mixing matrices ensure
$\mathrm{Im}~\tilde {J}^{ij}_{\alpha\beta} = \mathrm{Im}~\tilde {\bar J}^{ij}_{\alpha\beta} = 0$.
Therefore, the first term quantifies CP and CPT violation,
whereas the second one is CP and T violating.

Looking at the behavior of each of them under T and CPT,
we find that the components
\begin{align}
	\label{eq:ACPT}
	\comp{CPT}_{\alpha\beta} &\equiv
	-4\sum_{j<i}\left[
		\mathrm{Re}~\tilde J^{ij}_{\alpha\beta} \sin^2\tilde\Delta_{ij}
		-\mathrm{Re}~\tilde {\bar J}^{ij}_{\alpha\beta} \sin^2\tilde{\bar \Delta}_{ij}
	\right]\,,\\
	\label{eq:AT}
	\comp{T}_{\alpha\beta} &\equiv
	-2\sum_{j<i}\left[
		\mathrm{Im}~\tilde J^{ij}_{\alpha\beta} \sin 2\tilde\Delta_{ij}
		-\mathrm{Im}~\tilde {\bar J}^{ij}_{\alpha\beta} \sin 2\tilde{\bar \Delta}_{ij}
	\right]
\end{align}
have definite parities on \emph{both} symmetries:
$\comp{CPT}_{\alpha\beta}$ is CPT-odd and T-invariant,
whereas $\comp{T}_{\alpha\beta}$ is CPT-invariant and T-odd.
Also, by construction, both of them are CP-odd.
In the sense of these symmetry principles,
we call these quantities the \emph{disentangled} components of $\asym{CP}_{\alpha\beta}$,
since the way in which each of them violates CP is fundamentally different.
The fact that CPT holds in vacuum means that
effects of \emph{genuine} CP violation affect the CP asymmetry as 
the T-odd component, so a non-vanishing measurement of $\comp{T}_{\alpha\beta}$
is a proof of CP violation in the lepton sector.
On the other hand, the CPT-violating and T-invariant matter effects,
taking into account in the Hamiltonian~(\ref{eq:Htilde}) via the real parameter $a$,
contribute to the CP asymmetry as $\comp{CPT}_{\alpha\beta}$.

Without explicit expressions for the masses and mixings,
the way in which this disentanglement translates into experimental measurements
is through the definite T parity of both components,
which ensures that the matter-induced component
---the T-invariant $\comp{CPT}_{\alpha\beta}$--- %~(\ref{eq:ACPT})---
is an even function of the baseline,
whereas the genuine component
---the T-odd $\comp{T}_{\alpha\beta}$--- %~(\ref{eq:AT})---
is an odd function 
%(notice the $\sin^2 L$ in the first one and the $\sin 2L$ in the second one).
(notice the different functions of Eqs.(\ref{eq:ACPT}, \ref{eq:AT}) in the oscillation phases, 
which are proportional to $L$ at a given energy).
Therefore, the existence of an $L$-odd term in 
a measurement of the CP asymmetry as a function of the baseline
would be a direct test of CP violation in the lepton sector.
Since this kind of measurement is not feasible,
we study how these symmetry behaviors affect the dependence of each component
on the parameters in the Hamiltonian~(\ref{eq:Htilde}).

\section{Signatures at Future Accelerator Experiments}
\label{sec:signatures}

The difference between neutrinos and antineutrinos in the Hamiltonian 
lies in the signs of the complex phase $\delta$ and the matter parameter $a$,
which are associated to T ($\delta \mapsto -\delta$) and CPT ($a\mapsto -a$) transformations.
The definite parities of both components imply that 
$\comp{CPT}_{\alpha\beta}$ must be an even function of $\delta$ (typically $\cos\delta$)
and an odd function of $a$, and so it vanishes in vacuum,
whereas $\comp{T}_{\alpha\beta}$ must be an even function of $a$
and an odd function of $\delta$, an so it vanishes in the absence of genuine CP violation.

We calculate approximated expressions for both components
at energies between the two MSW resonances~\cite{Wolfenstein:1977ue,Mikheev:1986gs}
$\dm_{21} \ll a \ll \abs{\dm_{31}}$,
for the case of the golden $\nu_\mu \to \nu_e$ transitions.
The definite parity in $a$ ensures that higher-order corrections will be quadratic,
so our perturbation parameters are
\[
	\frac{\Delta m^2_{21}}{\Delta m^2_{31}} \sim 0.030\,,
	\hspace{0.5cm}
	\abs{U_{e3}}^2 \sim 0.022\,,
\]
\begin{equation}
	\left[\frac{\Delta m^2_{21}}{a}\right]^2 \!\sim \frac{0.12}{(E/\mathrm{GeV})^2}\,,
	\hspace{0.3cm}
	\left[\frac{a}{\Delta m^2_{31}}\right]^2 \!\sim 0.008\, (E/\mathrm{GeV})^2\,,
	\hspace{0.3cm}
	\left[\frac{aL}{4E}\right]^2 \!\sim 0.084 \left(\frac{L}{1000\mathrm{km}}\right)^2\,,
\end{equation}
where we used the mean value of the Earth mantle density~\cite{Mocioiu:2000st} in $a$.

\begin{figure}[t]
	\centering
	\begin{minipage}{0.485\textwidth}
		\begin{tikzpicture}[line width=1 pt, scale=1.5]
			\node at (0,0)[below left]{\includegraphics[width=\textwidth]{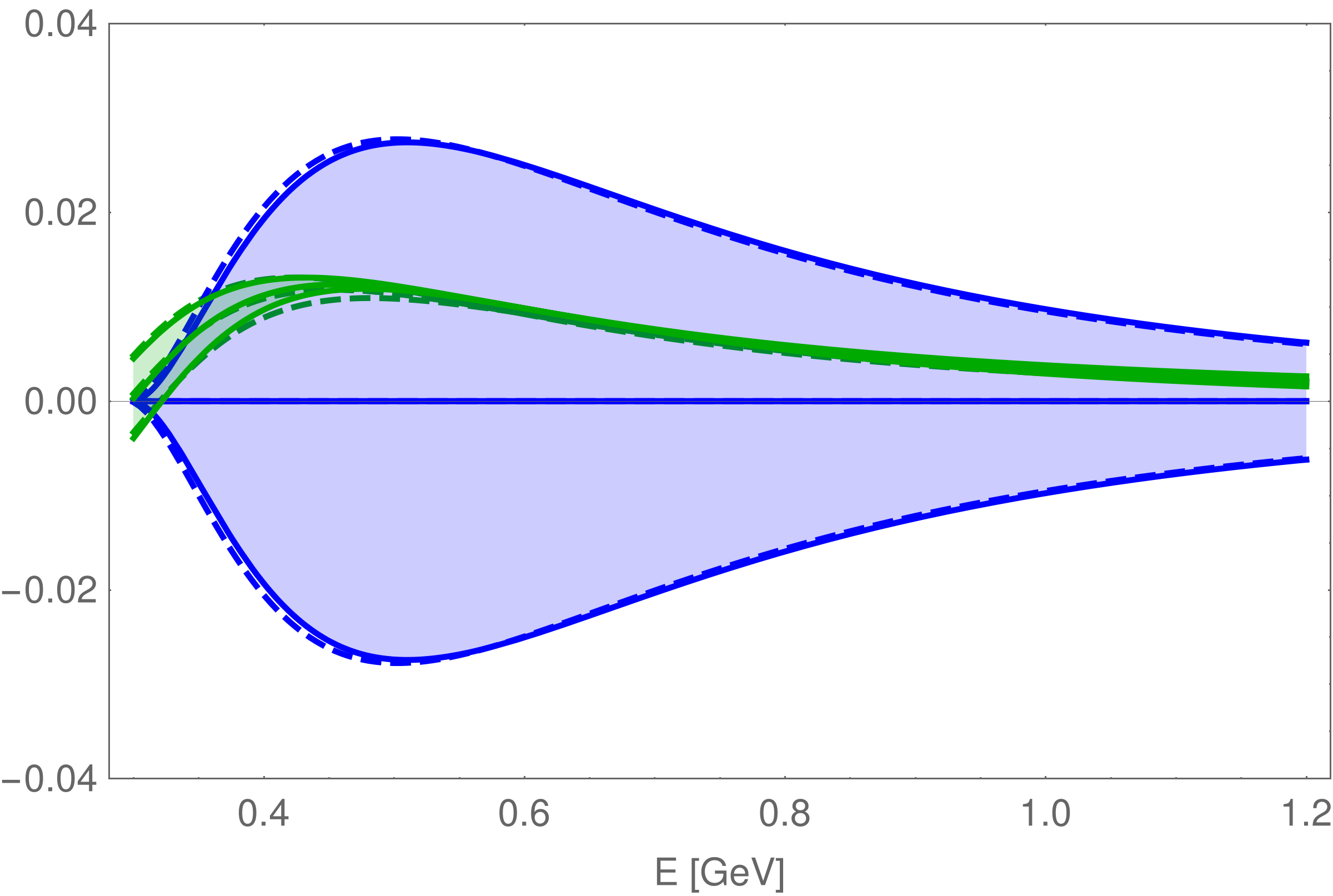}};		
			\node at (-0.1,-0.20)[below left]{$L = 295$~km};
			%\node at (-4.85,-0.20)[below right]{NH};
		\end{tikzpicture}
		%\caption{$L=295$~km}
	\end{minipage}
	\hfill
	\begin{minipage}{0.485\textwidth}
		\begin{tikzpicture}[line width=1 pt, scale=1.5]
			\node at (0,0)[below left]{\includegraphics[width=\textwidth]{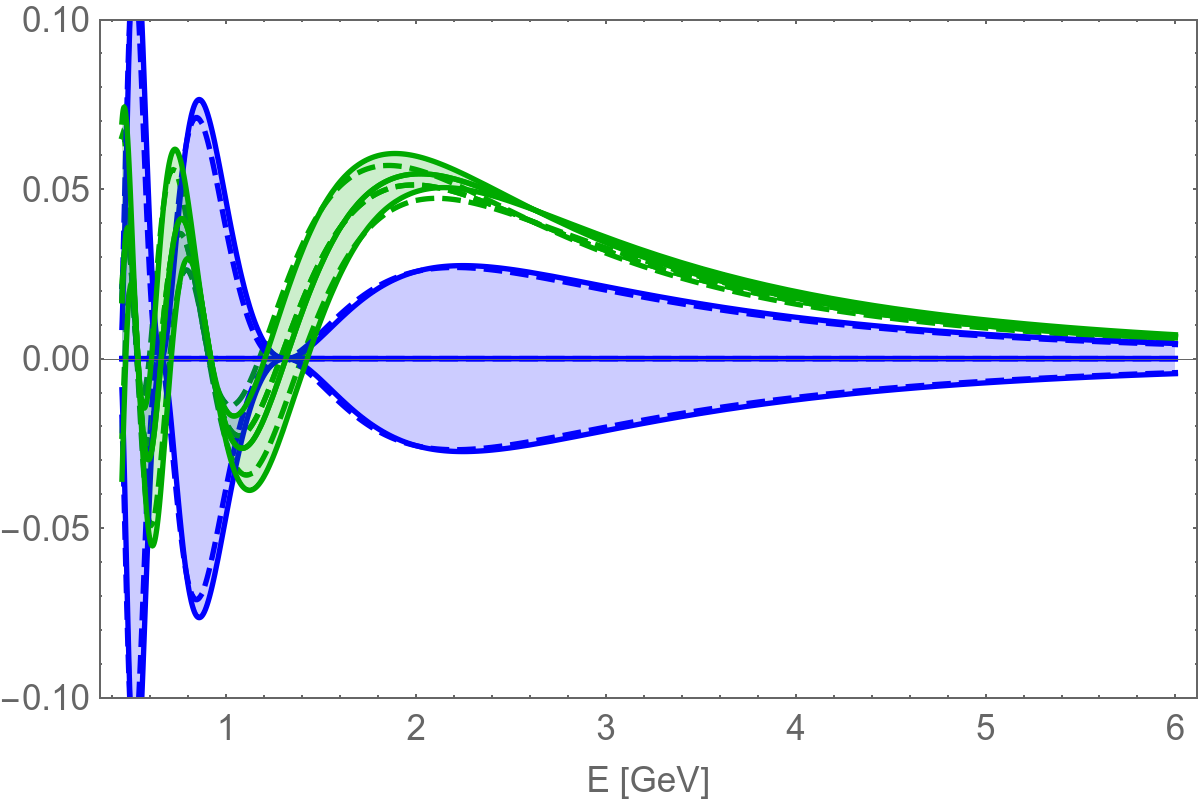}};		
			\node at (-0.1,-0.20)[below left]{$L = 1300$~km};
			%\node at (-4.7,-0.20)[below right]{NH};
		\end{tikzpicture}
		%\caption{$L=1300$~km}
	\end{minipage}
	\caption{
		$\comp{CPT}_{\mu e}$ (green) and $\comp{T}_{\mu e}$ (blue) 
		at T2HK (left) and DUNE (right) baselines.
		Analytical expression from Eqs.~(\ref{eq:anACPT}, \ref{eq:anAT}) (solid)
		and numerical results (dashed) shown.
		The bands show all possible values of the components
		changing $\delta$ in ($0,\, 2\pi$).
	}
	\label{fig:bands}
\end{figure}

We perturbatively solve for the eigenvalues ($\tilde M^2$) and eigenstates ($\tilde U$)
of the Hamiltonian~(\ref{eq:Htilde}) assuming constant matter density,
which lead to the approximated expressions for the disentangled components
\begin{align}
	\label{eq:anACPT}
	\comp{CPT}_{\mu e} &=
           16\, \Delta_a
           \left[ \frac{\sin\Delta_{31}}{\Delta_{31}}-\cos\Delta_{31} \right]
           \left( S\sin\Delta_{31} + J_r\cos\delta\, \Delta_{21}\cos\Delta_{31} \right)
           +\mathcal{O}(\Delta_a^3)\,,\\
	\label{eq:anAT}
	\comp{T}_{\mu e} &=
           -16\,J_r \sin\delta\, \Delta_{21}\sin^2\Delta_{31}
           +\mathcal{O}(\Delta_a^2) \,,
\end{align}
where $S \equiv c_{13}^2 s_{13}^2 s_{23}^2$,
$J_r \equiv c_{12} c_{13}^2 c_{23} s_{12} s_{13} s_{23}$,
$\Delta_a\equiv \frac{aL}{4E} \propto L$ and
$\Delta_{ij} \equiv \frac{\Delta m^2_{ij} L}{4E} \propto~\!L/E$.
These expressions are shown together with 
the exact (numerical) diagonalization of the Hamiltonian in Figure~\ref{fig:bands}.
They have errors at the few percent level, but the overall behavior is well reproduced.
The key feature is the understanding of the position of the zeros of $\comp{CPT}_{\mu e}$,
which are very well reproduced,
since those are the points where the CP asymmetry itself is free from matter effects.
Therefore, these precise-enough expressions are a good tool to understand the
peculiar signatures of each component.

\subsection{Hierarchy discrimination}

We find that, at leading order, 
the genuine component $\comp{T}_{\mu e}$~(\ref{eq:anAT}) is Hierarchy-independent,
i.e. independent of sign($\dm_{31}$),
so all information in the CP asymmetry about the mass ordering 
must come from the matter-induced component $\comp{CPT}_{\mu e}$~(\ref{eq:anACPT}).
This component has no definite parity under a change of Hierarchy, 
so we look separately at its Hierarchy-odd $\delta$-independent term and 
its Hierarchy-independent $\delta$-dependent term,
\begin{align}
	\label{eq:ACPT-}
	\comp{CPT}_- &\equiv
           16\, \Delta_a
           \left[ \frac{\sin\Delta_{31}}{\Delta_{31}}-\cos\Delta_{31} \right]
           S\sin\Delta_{31} \,,\\
	\label{eq:ACPT+}
	\comp{CPT}_+ &\equiv
           16\, \Delta_a
           \left[ \frac{\sin\Delta_{31}}{\Delta_{31}}-\cos\Delta_{31} \right]
           J_r\cos\delta\, \Delta_{21}\cos\Delta_{31} \,.
\end{align}

The fact that the $\delta$-dependent term $\comp{CPT}_+$ is proportional to $\Delta_{21}$,
i.e. to $1/E$, means that it is negligible at high enough energy,
ensuring that the whole matter-induced component $\comp{CPT}_{\mu e}$ is Hierarchy-odd.
Our expressions show that the condition $\abs{\comp{CPT}_-} > \abs{\comp{CPT}_+}$ holds
for energies $E > 1.1\, E_{1^\mathrm{st}\text{ node}}$ above the first node of the vacuum oscillation.
At the T2HK baseline, this includes the whole energy spectrum;
however, as seen in Fig.~\ref{fig:bands}, 
the dominance of the genuine component over the matter-induced one
forbids the exploitation of this fact in a measurement of the observable CP asymmetry
to test the Hierarchy.

At DUNE baseline, the matter-induced component is Hierarchy-odd at energies above
$1.4$~GeV, which correspond to the region in which it is larger than the genuine component.
The measurement of the (sign of the) CP asymmetry at any of these points 
would thus determine the Hierarchy.

%\begin{figure}[t]
%	\centering
%	\begin{minipage}{0.485\textwidth}
%		\begin{tikzpicture}[line width=1 pt, scale=1.5]
%			\node at (0,0)[below left]{\includegraphics[width=\textwidth]{img/ACPTpmHK.png}};
%			\node at (-0.15,-0.20)[below left]{$L = 295$~km};
%		\end{tikzpicture}
%		%\caption{$L=295$~km}
%	\end{minipage}
%	\hfill
%	\begin{minipage}{0.485\textwidth}
%		\begin{tikzpicture}[line width=1 pt, scale=1.5]
%			\node at (0,0)[below left]{\includegraphics[width=\textwidth]{img/ACPTpmDUNE.png}};		
%			\node at (-0.1,-0.20)[below left]{$L = 1300$~km};
%		\end{tikzpicture}
%		%\caption{$L=1300$~km}
%	\end{minipage}
%	\caption{
%	}
%\end{figure}

\subsection{Direct observation of CP violation}

The measurement of a non-vanishing value of the genuine component $\comp{T}_{\mu e}$~(\ref{eq:anAT})
is a proof of CP violation in the lepton sector.
To extract this information from the observable CP asymmetry,
we suggest different strategies depending on the baseline.

At the medium baseline of T2HK, the matter-induced component is smaller than the genuine one.
It is also hierarchy-odd and nearly $\delta$-independent. 
Therefore, once the Hierarchy is measured, this component can be subtracted
from the experimental CP asymmetry as a theoretical background
in order to obtain the value of the genuine component.

Due to its longer baseline, the matter-induced component dominates the CP asymmetry at DUNE in general.
A measurement of the genuine component is thus only possible at those points where the matter-induced
one vanishes.
From the expression~(\ref{eq:anACPT}) for $\comp{CPT}_{\mu e}$ we find a family of
$\delta$-independent zeros given by
\begin{equation}
	\label{eq:MP}
	\tan\Delta_{31} = \Delta_{31}\,,
\end{equation}
at energies close to the oscillation maxima $\sin^2\Delta_{31}=1$
and the maximal values of the genuine component
\begin{equation}
	\tan\Delta_{31} = -2\Delta_{31}\,.
\end{equation}

\begin{figure}[t]
	\mbox{}
	\hfill
	\begin{minipage}{0.45\textwidth}
		\begin{tikzpicture}[line width=1 pt, scale=1.5]
			\node at (0,0)[below left]{\includegraphics[width=\textwidth]{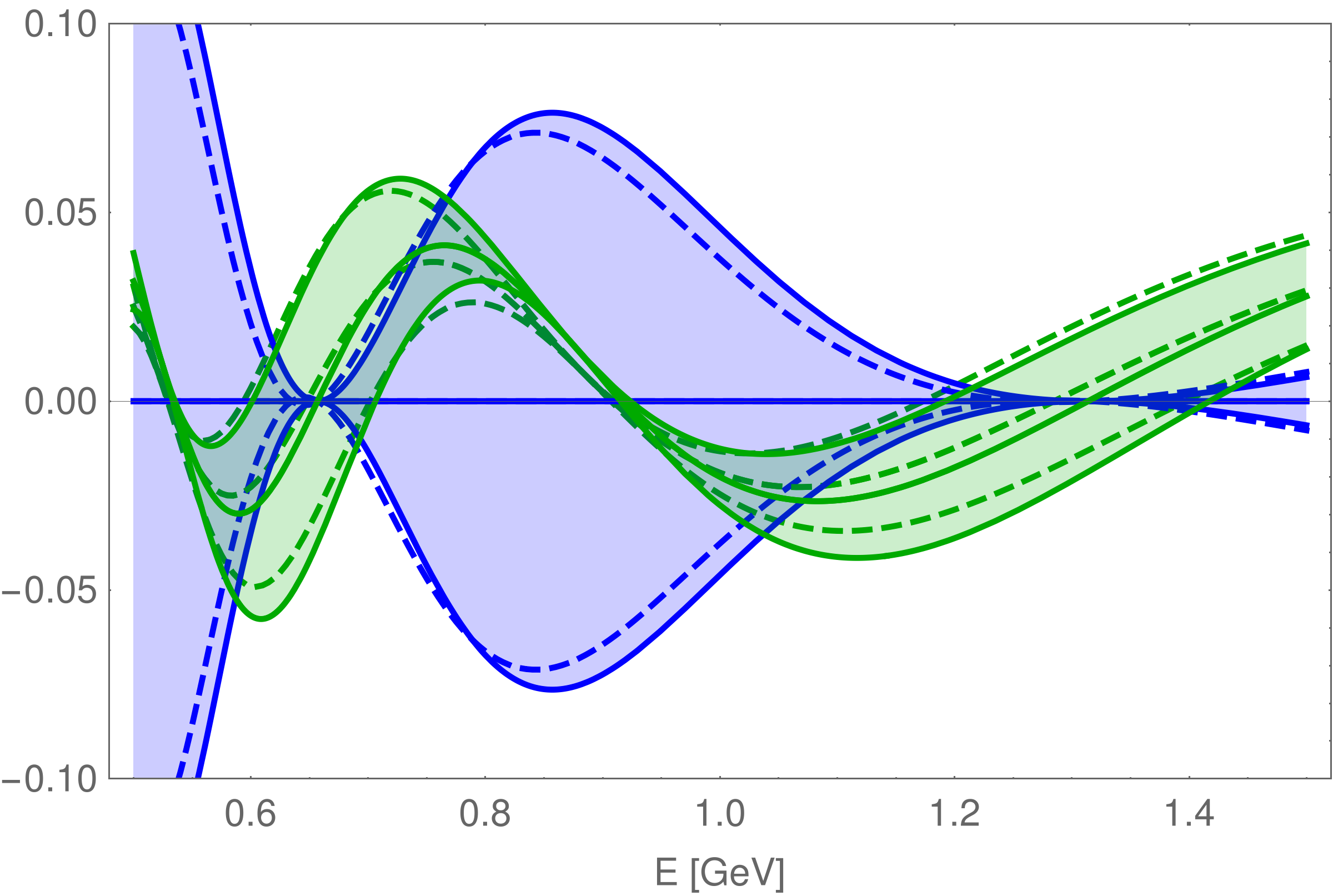}};
			%\node at (-0.1,-0.20)[below left]{NH};
		\end{tikzpicture}
		%\caption{$L=295$~km}
	\end{minipage}
	\hfill
	\begin{minipage}{0.45\textwidth}
		\begin{tikzpicture}[line width=1 pt, scale=1.5]
			\node at (0,0)[below left]{\includegraphics[width=\textwidth]{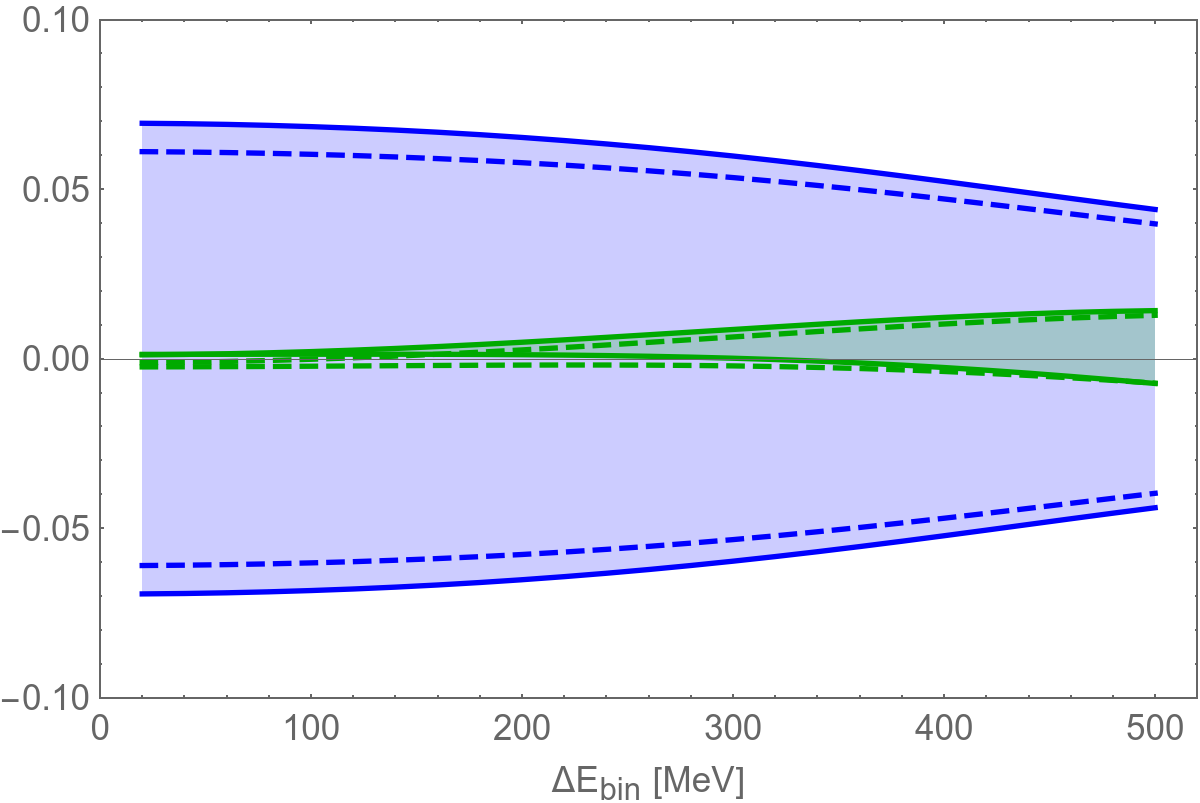}};
			%\node at (-0.1,-0.20)[below left]{NH};
		\end{tikzpicture}
		%\caption{$L=295$~km}
	\end{minipage}
	\hfill
	\mbox{}
	\caption{
		\texttt{Left:}
		zoom of Figure~\ref{fig:bands} at DUNE baseline.
		\texttt{Right:}
		Averaged value of the disentangled components around
		$E = 0.92$~GeV in an energy bin width $\Delta E_\mathrm{bin}$.
		\vspace{-3pt}
	}
	\label{fig:zoom}
\end{figure}

The highest-energy solution of this family of zeros happens at $L/E = 1420$~km/GeV,
i.e. $E=0.92$~GeV at DUNE baseline.
Notice that this result is Hierarchy-independent, as well as independent of
all mixing angles; it comes from the condition~(\ref{eq:MP}) for $\Delta_{31}$,
so it only depends on the value of $\abs{\dm_{31}}$.
We show in Figure~\ref{fig:zoom} (left) a zoom of the DUNE spectrum in Figure~\ref{fig:bands}
around this energy,
where one can clearly see that the zero of the matter-induced component $\comp{CPT}_{\mu e}$
is $\delta$-independent (the bands collapse to a line) and close to a maximal value
of the genuine component $\comp{T}_{\mu e}$.

The matter-induced component changes sign around the zero,
so the average value of the component in a finite energy bin around the zero
is still free from matter effects.
Since the genuine component is close to maximal, 
this averaging has little effect on its value.
We show these properties in Figure~\ref{fig:zoom} (right),
from which we find that a measurement of the CP asymmetry $\asym{CP}_{\mu e}$
around $E=0.92$~GeV at DUNE baseline, with a bind width up to $200$~MeV
is free from matter effects:
it is as clean a test of genuine CP violation as it would be in vacuum.

%$\begin{figure}[t]
%$	\hfill
%$	\includegraphics[width=0.5\textwidth]{img/tanx.png}
%$	\hfill
%$	\begin{minipage}[b]{0.45\textwidth}
%$		\caption{
%$			Illustration of the analytical conditions for
%$			a vanishing $\comp{CPT}_{\mu e}$~() 
%$			and maximal $\comp{T}_{\mu e}$~().
%$			The red ellipse emphasizes the highest-energy solution.
%$			The orange region shows the possible position of the $\delta$-dependent
%$			zeros of $\comp{CPT}_{\mu e}$~().
%$		}
%$	\end{minipage}
%$\end{figure}

%\begin{figure}[t]
%	\centering
%	\begin{minipage}{0.485\textwidth}
%		\begin{tikzpicture}[line width=1 pt, scale=1.5]
%			\node at (0,0)[below left]{\includegraphics[width=\textwidth]{img/bins.png}};
%			\node at (-0.1,-0.20)[below left]{NH};
%		\end{tikzpicture}
%		%\caption{$L=295$~km}
%	\end{minipage}
%	\hfill
%	\begin{minipage}{0.485\textwidth}
%		\begin{tikzpicture}[line width=1 pt, scale=1.5]
%			\node at (0,0)[below left]{\includegraphics[width=\textwidth]{img/bins_IH.png}};
%			\node at (-0.1,-0.20)[below left]{IH};
%		\end{tikzpicture}
%		%\caption{$L=1300$~km}
%	\end{minipage}
%	\caption{
%	}
%\end{figure}

\section{Conclusions}
\label{sec:conclusions}

We present a Disentanglement Theorem able to separate, from symmetry principles,
the genuine and matter-induced components of the CP asymmetry $\asym{CP}_{\alpha\beta}$.
From the definite parities of these components under T and CPT
we find that the genuine component $\comp{T}_{\alpha\beta}$
is an even function of the matter parameter $a$ and an odd function of $L$ and $\delta$,
whereas the matter-induced component $\comp{CPT}_{\alpha\beta}$
is odd in $a$ and even in $L$ and $\delta$.

We exploit these parities to build approximate analytical expressions of these components
in the $\nu_\mu \to \nu_e$ flavor channel.
A non-vanishing genuine component tests CP violation in the lepton sector,
whereas only the matter-induced component changes sign under a change of Hierarchy.
This Hierarchy dependence allows us to show only plots for Normal Hierarchy;
the case of Inverted Hierarchy would show the same
genuine component and a matter-induced component with opposite sign.
We analyze the signatures of the disentangled components
at the baselines of T2HK and DUNE.

At the medium baseline of T2HK, the matter-induced component is smaller than the genuine one.
At the $L/E$ reachable by the experiment,
$\comp{CPT}_{\mu e}$ is nearly $\delta$-independent,
so it can be subtracted from the experimental CP asymmetry 
provided the hierarchy is known.
This would allow to measure the genuine component
$\comp{T}_{\mu e} = \asym{CP}_{\mu e} - \comp{CPT}_{\mu e}$.

At the longer baseline of DUNE, both the Hierarchy and the genuine component are accessible.
The whole CP asymmetry changes its sign under a change of Hierarchy at energies above $1.4$~GeV,
so it can be used to measure the mass ordering.
On the other hand,
the matter-induced component vanishes $\delta$-independently at
%$E = 0.92~\mathrm{GeV}\, (L/1300~\mathrm{km})$,
\begin{equation}
	E = 0.92~\mathrm{GeV}\, 
	\frac{L}{1300~\mathrm{km}}\,
	\frac{\abs{\dm_{31}}}{2.5\times 10^{-3}~\mathrm{eV}^2}\,,
\end{equation}
so the measurement of the CP asymmetry at this point is 
a direct test of CP violation in the lepton sector,
uncontaminated by matter effects.

\vspace{-6pt}
\ack

This research has been supported by 
MINECO Project FPA 2017-84543-P, 
Generalitat Valenciana Project GV PROMETEO 2017-033
and
Severo Ochoa Excellence Centre Project SEV 2014-0398.
A.S. acknowledges the MECD support through the FPU14/04678 grant.

\section*{References}
\bibliography{refs}

\end{document}